\documentstyle[12pt]{article}
\def\spose#1{\hbox to 0pt{#1\hss}}
\def\lta{\mathrel{\spose{\lower 3pt\hbox{$\mathchar"218$}}
     \raise 2.0pt\hbox{$\mathchar"13C$}}}
\def\gta{\mathrel{\spose{\lower 3pt\hbox{$\mathchar"218$}}
     \raise 2.0pt\hbox{$\mathchar"13E$}}}
\def\n{\noindent}
\def\cl{\centerline}

\def\be{\begin{equation}}
\def\ee{\end{equation}}
\def\msun{M_{\odot}}

\def\mdot{\dot M}

\def\refs{\noindent \hangindent 20pt }
\def\aaa#1{{A\&A,} {#1}}

\def\apj#1{{ApJ,} {#1}}
\def\apjs#1{{ApJS,} {#1}}
\def\mnras#1{{MNRAS,} {#1}}

\begin{document}
\baselineskip=24pt

~~\vskip 1.0in
\cl {\large \bf On the morphology of accretion flows}
\vskip 0.2in

\cl {\large \bf with small non-zero specific angular momentum}

\vskip 0.5in

\cl {Xingming Chen\footnote[1]{Department of Physics \& Astronomy, 
Northwestern University, Evanston, IL 60208}$^{,2}$, Ronald E. Taam$^{1}$, 
Marek A. Abramowicz\footnote[2]
{Department of Astronomy \&
Astrophysics, G{\"o}teborg University and Chalmers University of Technology, 
412 96 G{\"o}teborg, Sweden}, and}

\vskip 0.2in

\cl {Igor V. Igumenshchev$^{2,}$\footnote[3]
{Institute of Astronomy, 48 Pyatnitskaya Street, Moscow,
117810, Russia}}

\vskip 2.0in

\n{Received:}

\newpage
\n{\bf Abstract}
\medskip

The morphology of adiabatic accretion flows with small non-zero specific 
angular momentum has been investigated in the axisymmetric and nonviscous
limit.  For an initial state characterized by a Bondi flow with the 
specific angular momentum distributed with respect to polar angle, a 
travelling shock wave forms which propagates more rapidly in the equatorial 
plane than in the plane perpendicular to it resulting in the formation of 
a hot torus.  In cases where the incoming flow is restricted to lie near the 
equatorial plane, a strong wind forms directed away from this plane with the 
tendency for the formation of a nonsteady shock structure.  As the height of 
the incoming flow is increased it is found that the resultant wind weakens. 
The parameter regime which delineates accretion flows characterized by a 
travelling shock and a nearly standing shock structure is presented. 

\vskip 0.2in

\n{\bf keywords:}
accretion, accretion disks --- hydrodynamics --- methods: numerical

\newpage
\n{\bf 1 Introduction}
\medskip

The process of accretion is central to  our understanding of many diverse 
astronomical objects and phenomena.  The formation of young stellar objects 
on the galactic scale and the underlying cause of the activity of  
galactic nuclei on the extragalactic scale are just a few such examples.  A 
realistic description of accretion flows is likely to be multi-dimensional and 
time dependent in character and may involve the interplay between various 
complex physical processes such as dissipative magnetohydrodynamics and 
radiation. However, in some idealized cases simple geometry or physics may 
suffice to facilitate solution by analytical or semi-analytical means. Among 
such problems include spherical accretion in the adiabatic or isothermal 
approximation, and disk accretion driven by viscous effects.

The classical case of spherical accretion onto a stationary point mass 
in the hydrodynamical approximation was
first considered by Bondi (1952). In this study he showed the existence  
of several families of solutions and described their topology.  Of import
is the transonic solution which bears his name.  The extension of the 
accretion problem to the two dimensional case in which the accretor is 
in relative motion with respect to the surrounding medium was actually 
investigated earlier in the particle approach by Hoyle \& Lyttleton (1939) 
and Bondi \& Hoyle (1944).  Later, 
Ruderman \& Spiegel (1971), Wolfson (1977) and Bisnovatyi-Kogan et al. 
(1979) reexamined this problem semi-analytically in the hydrodynamical 
approximation.  However, no analytical 
solution exists in the general case in which the hydrodynamical effects 
are fully considered.  In fact, the recent numerical calculations reveal 
solutions which are complex and time dependent. 
In particular, the flows are found to  
exhibit a wide range of structures including shocks, jets, and  
transient accretion disks
(see, e.g., Shima et al. 1985; Taam \& Fryxell 1988; Ruffert \& Arnett 1994).

Accretion flows with large specific angular momentum have also been the 
subject of intensive theoretical 
study following the seminal analytical work of Shakura \& Sunyaev (1973).  
In the case where matter is viscously driven recent numerical studies have
focused on the formation of a disk wind (Eggum, Coroniti, \& Katz 
1987, 1988) and meridional flows in geometrically thin accretion disks
(see Kley \& Lin 1992; R{\'o}{\.z}yczka, Bodenheimer, \& Bell 1994). In 
such studies viscosity and radiation transfer are essential for a proper
description of the disk. 
On the other hand, if the specific angular momentum of the accreting matter is 
low, then the flow is, in general, pressure driven and the disk is 
geometrically thick.  
Accretion in this regime was studied in the pioneering investigations by 
Hawley, Smarr, \& Wilson (1984a,b) and Clarke, Karpik \& Henriksen (1985),
and recently by Molteni, Lanzafame, \& Chakrabarti (1994), Ryu et al (1995), 
and Igumenshchev, Chen, \& Abramowicz (1996).  The appearance of vortices
and shock waves are common features in these accretion flows. 

The results from such investigations, taken as an aggregate, do not present
a clear picture of disk accretion in the geometrically thick disk limit.
For example, Hawley et al. (1984b) studied the time dependent two dimensional
flows in the vicinity of a black hole in the limit that the specific angular 
momentum of the accreting matter lies between that of the marginally stable 
orbit and the marginally bound orbit.  In this case, a shock front forms and 
travels outward in the disk.  This result was confirmed in an independent 
study by Clarke et al. (1985) in the Newtonian approximation.  Their 
results, however, are in direct contrast to the results obtained in the recent 
studies of Molteni et al. (1994) and Ryu et al. (1995). In both these latter 
investigations, a standing shock front formed in the disk confirming the 
earlier work of Chakrabarti (1990) and Chakrabarti \& Molteni (1993).  
Furthermore, both studies demonstrated that a wind emanating from the inner 
disk region is formed.  Further complicating the description of the accretion 
process is the recent study by Igumenshchev et al. (1996) who investigated 
similar thick disks, but with viscous effects included. In this case shock 
formation was absent in the accretion flow in conflict with the conclusion 
reached in Molteni et al. (1994). 

In this paper we report on the results of two dimensional hydrodynamical 
simulations which delineate the parameter regime where accretion flows are
characterized by a travelling shock front and a standing shock front in the
inner disk.  To make comparisons more meaningful to previous studies, we   
restrict our study to accretion flows with
small non-zero specific angular momentum in the adiabatic, nonviscous
hydrodynamical approximation.   Our goal is to provide for an understanding
of the various accretion flows in terms of the vertical extent, specific
angular momentum of the incoming matter (see Hawley et al. 1984a,b), and the 
boundary conditions imposed on the flow.  In \S 2 we outline the specific 
approximations underlying our investigation.  The initial conditions and the 
boundary conditions for the suite of problems is outlined in \S 3 and their 
numerical results are presented in \S 4.  Finally, we discuss our results in 
the context of previous studies in the last section.

\vskip 0.5in
\n {\bf 2 Formulation}
\vskip 0.3in

In this investigation it is most convenient to use spherical coordinates 
$(r, \theta,\varphi)$. The accretion disk is assumed to be non 
self-gravitating 
and axisymmetric with the rotation axis coincident with the polar axis ($\theta
=0$).  The flow is assumed to be adiabatic with no local radiative cooling 
included.  The gravitational field of the black hole is described in terms of 
a Newtonian potential in order to compare to previous studies 
$$\Phi = -{GM\over r} \eqno(2.1) $$
where $M$ the mass of the central object.

The governing equations describing the flow in these coordinates can be 
expressed in the following form.  For the equation of continuity, 
%%%%%%%%%%%%%%%%%%%%%%%%%%%%%%%%%%%%%%%%%%%%%%%%%%%%%%%%%%%%%%%%%%%%%%%%%%%%
$$ {\partial\rho \over \partial t}+{1\over r^2}{\partial\over\partial r}
   (r^2\rho v_r) + {1 \over r \sin \theta}{\partial\over\partial \theta}
   (\rho v_\theta \sin \theta) = 0. \eqno(2.2) $$
%%%%%%%%%%%%%%%%%%%%%%%%%%%%%%%%%%%%%%%%%%%%%%%%%%%%%%%%%%%%%%%%%%%%%%%%%%%%

\n The equations of motion take the form
%%%%%%%%%%%%%%%%%%%%%%%%%%%%%%%%%%%%%%%%%%%%%%%%%%%%%%%%%%%%%%%%%%%%%%%%%%%%
$$ {\partial\over\partial t}(\rho v_r) + {1\over r^2}{\partial\over\partial r}
   (r^2 \rho v_r^2) + {1 \over r \sin \theta}
   {\partial\over\partial \theta}(\rho v_r v_\theta \sin \theta) =
   -\rho {\partial \Phi \over \partial r} - {\partial p \over\partial r}
   +{\rho({v_\theta}^2+{v_\varphi}^2) \over r}
%$$
%$$
%   +{1\over r^2}{\partial\over\partial r}
%   (r^2 \tau_{rr}) + {1 \over r \sin \theta}
%   {\partial\over\partial \theta}(\tau_{r\theta} \sin \theta)
%   -{\tau_{\theta \theta}+\tau_{\varphi \varphi} \over r}
   , \eqno(2.3)$$
%%%%%%%%%%%%%%%%%%%%%%%%%%%%%%%%%%%%%%%%%%%%%%%%%%%%%%%%%%%%%%%%%%%%%%%%%%%%

$$ {\partial\over\partial t}(\rho v_\theta) + {1\over r^2}
   {\partial\over\partial r}(r^2\rho v_\theta v_r) + {1 \over r \sin \theta}
   {\partial\over\partial \theta}(\rho v_\theta^2 \sin \theta) =
   -{1 \over r}{\partial p \over\partial \theta} 
   -{\rho v_\theta v_r \over r} + {\rho v_\varphi^2 \cot \theta \over r}
%$$
%$$
%   +{1\over r^2}{\partial\over\partial r}
%   (r^2 \tau_{\theta r}) + {1 \over r \sin \theta}
%   {\partial\over\partial \theta}(\tau_{\theta \theta} \sin \theta)
%   +{\tau_{r \theta} \over r}
%   -{\tau_{\varphi \varphi} \cot \theta \over r}
   , \eqno(2.4)$$
%%%%%%%%%%%%%%%%%%%%%%%%%%%%%%%%%%%%%%%%%%%%%%%%%%%%%%%%%%%%%%%%%%%%%%%%%%%%

$$ {\partial\over\partial t}(\rho \ell) + {1\over r^2}
   {\partial\over\partial r}(r^2\rho \ell v_r) + {1 \over r \sin \theta}
   {\partial\over\partial \theta}(\rho \ell v_\theta \sin \theta) = 0.
%   r \sin \theta 
%$$
%$$
%\left [
%   +{1\over r^2}{\partial\over\partial r}
%   (r^2 \tau_{\varphi r}) + {1 \over r \sin \theta}
%   {\partial\over\partial \theta}(\tau_{\varphi \theta} \sin \theta)
%   +{\tau_{r \varphi} \over r}
%   +{\tau_{\theta \varphi} \cot \theta \over r}
%\right]
   \eqno(2.5)$$
%%
%$$ {\partial\over\partial t}(\rho v_\varphi) + {1\over r^2}
%   {\partial\over\partial r}(r^2\rho v_\varphi v_r) + {1 \over r \sin \theta}
%   {\partial\over\partial \theta}(\rho v_\varphi v_\theta \sin \theta) =-{
%\rho v_\varphi v_r \over r} - {\rho v_\varphi v_\theta \cot \theta \over r}$$
%$$
%%%%%%%%%%%%%%%%%%%%%%%%%%%%%%%%%%%%%%%%%%%%%%%%%%%%%%%%%%%%%%%%%%%%%%%%%%%%

\n The energy equation is written as 
%%%%%%%%%%%%%%%%%%%%%%%%%%%%%%%%%%%%%%%%%%%%%%%%%%%%%%%%%%%%%%%%%%%%%%%%%%%%
$$ {\partial\over\partial t}(\rho\varepsilon) +
   {1\over r^2}{\partial\over\partial r}(r^2\rho\varepsilon v_r)
   +{1 \over r \sin \theta}
   {\partial\over\partial \theta}(\rho \varepsilon v_\theta \sin \theta)=
   - p\,{\rm div}\,\vec{v} 
%   + Q
   ,\eqno(2.6)$$
%%%%%%%%%%%%%%%%%%%%%%%%%%%%%%%%%%%%%%%%%%%%%%%%%%%%%%%%%%%%%%%%%%%%%%%%%%%%

\n with the equation of state taken to be,
%$$ p = {R\over\mu}\rho T + {1\over 3} a T^4 , \eqno(7')$$
%$$ \rho\varepsilon = {3\over 2}{R\over\mu}\rho T + a T^4 . \eqno(7'')$$
%%%%%%%%%%%%%%%%%%%%%%%%%%%%%%%%%%%%%%%%%%%%%%%%%%%%%%%%%%%%%%%%%%%%%%%%%%%%
$$ p=(\gamma-1)\rho\varepsilon. \eqno(2.7)$$
%%%%%%%%%%%%%%%%%%%%%%%%%%%%%%%%%%%%%%%%%%%%%%%%%%%%%%%%%%%%%%%%%%%%%%%%%%%%
Here $\rho$, $(v_r, v_\theta, v_\varphi)$, $\varepsilon$, and $p$
are the density, the velocities, the specific internal energy, and the pressure
of the flow respectively.  In addition, $\ell=v_\varphi r \sin \theta$ is the 
specific angular momentum, and $\gamma$ is the adiabatic index assumed
to be a constant ($\gamma=1.5$ or $5/3$). 

For definiteness we have fixed $M=10 \msun$ and $\ell(r_{out},\pi/2)
=\ell_0=2R_G c$
and have used $R_G$ and $R_G/c$ for the units of length and time respectively.
Here, $R_G = 2 GM/c^2$, is the Schwarzschild radius for a black hole of mass
$M$. In these units, the velocity is expressed in terms of the speed of 
light, $c$, and the energy per unit mass
in terms of $c^2$.  For variations in $M$ and $\ell_0$, the results 
can be scaled by changing the units of length and time to 
$(\ell_0/2R_Gc)^2 R_G$ and $(\ell_0/2R_G c)^3 R_G/c$ respectively.

The above equations are solved using the explicit second-order Eulerian 
hydrodynamical PPM method (Woodward \& Colella 1984). 
In the present study, the 
calculational domain is located within $r_{in} \leq r \leq r_{out}$ and 
$0 \leq \theta \leq \pi/2$, where $r_{in}$ and $r_{out}$ define the 
inner and outer 
boundaries respectively.  The numerical grid in the $\theta$ direction is 
homogeneous, while in the radial direction, the width of the radial zone is 
given by $\triangle r= r 
\triangle \theta/ (1-0.5 \triangle \theta)$.  We assume axial symmetry with 
respect to the polar axis $\theta=0$ and mirror symmetry with respect to the 
equatorial plane $\theta=\pi/2$.  At the inner radial boundary, 
zero derivatives 
with respect to radius $r$ are used for variables such as $v_\theta$ and 
$v_\varphi$, and linear extrapolations are used for $\rho$, $v_r$, and 
$\varepsilon$.  At the outer boundary, the values of all the variables are 
fixed in time. 

\vskip 0.5in
\n{\bf 3 Initial Conditions}
\vskip 0.3in

For a polytropic equation of state, $p=K\rho^\gamma $ where 
$K$ is the polytropic constant, the Bernoulli function of the flow becomes
%%%%%%%%%%%%%%%%%%%%%%%%%%%%%%%%%%%%%%%%%%%%%%%%%%%%%%%%%%%%%%%%%%%%%%%%%%%%
$${\cal B} = {\gamma \over\gamma-1}K \rho^{(\gamma-1)} + {v_r^2 \over 2}
+ {v_\theta^2 \over 2} + {l^2 \over 2 (r\sin \theta)^2} + 
\Phi. \eqno(3.1) $$
%%%%%%%%%%%%%%%%%%%%%%%%%%%%%%%%%%%%%%%%%%%%%%%%%%%%%%%%%%%%%%%%%%%%%%%%%%%%
The initial state of the flow can be constructed under additional assumptions.

\vskip 0.5in
\n{\bf 3.1 Bondi type conditions}
\vskip 0.3in

To construct a Bondi type condition, we first assume $v_\theta=0$ and $\ell=0$.
Then equations (3.1) and (2.2) become respectively,
%%%%%%%%%%%%%%%%%%%%%%%%%%%%%%%%%%%%%%%%%%%%%%%%%%%%%%%%%%%%%%%%%%%%%%%%%%%%
$$ {\gamma \over\gamma-1} K \rho^{\gamma-1} + {v_r^2 \over 2}
- {GM \over r} ={\cal B}=const, \eqno(3.2) $$
%%%%%%%%%%%%%%%%%%%%%%%%%%%%%%%%%%%%%%%%%%%%%%%%%%%%%%%%%%%%%%%%%%%%%%%%%%%%
and
%%%%%%%%%%%%%%%%%%%%%%%%%%%%%%%%%%%%%%%%%%%%%%%%%%%%%%%%%%%%%%%%%%%%%%%%%%%%
$$ 4 \pi r^2 \rho v_r = - \mdot  , \eqno(3.3) $$
%%%%%%%%%%%%%%%%%%%%%%%%%%%%%%%%%%%%%%%%%%%%%%%%%%%%%%%%%%%%%%%%%%%%%%%%%%%%
where $\mdot$ is the mass accretion rate.
It is common to assume the sound speed at infinity as $c_\infty$, which is
related to the Bernoulli parameter as ${\cal B} =c_\infty^2/(\gamma-1)$,
where the local adiabatic sound speed is
$c_\infty^2 = \gamma p_\infty/\rho_\infty = K \gamma \rho_\infty^{(\gamma-1)}$. 
With $c_\infty$ one can define the Bondi radius as $R_B=GM/c_\infty^2$.
thus, ${\cal B}/c^2={1 \over 2(\gamma -1)} {R_G \over R_B}$.
For the transonic solution which we choose, specification of only two of the 
three constants, ${\cal B}$, $\mdot$, and $K$ is required because
the regularity condition at the sonic point provides a constraint among 
the constants which gives the maximal accretion rate. For $\gamma = 5/3$ the 
sonic radius is at the origin.  To consider a more general case, we use 
$\gamma=1.5$.

In the case where the specific angular momentum does not vanish, we assume 
that the initial condition can be described by a superposition of the specific
angular momentum on the Bondi solution.  Assuming that $v_\theta=0$ and the 
Mach number, the polytropic constant, and the Bernoulli function of the flow 
remain unchanged, we take the radial velocity and 
the density to be determined according to equation (3.2).  Two cases for the 
distribution of specific angular momentum are considered.  In the first case
$$\ell(r,\theta)=\ell_0 (1-\cos \theta), \eqno(3.4)$$
and the other is described by  
$$\ell(r,\theta)=\ell_0=const. \eqno(3.5)$$
It should be noted that, in the latter case, when $\theta$ is small, 
no solution may exist since $\ell$ does not vanish. 
Under such a situation, $v_r$ is set to zero and 
a numerically small quantity is assigned to $\rho$.

We have also considered cases in which the flow is restricted near the
equatorial plane. In these cases, a vertical extent of the flow, $H$, is 
required.  For $r\cos \theta > H$, the flow is truncated, i.e., the velocities 
are set to zero and a numerically small value is assigned to $\rho$.

\vskip 0.5in
\n{\bf 3.2 Parallel type conditions}
\vskip 0.3in

To compare with the previous study of Ryu et al. (1995), we consider parallel 
type initial conditions in which the velocity of the incoming flow is parallel 
to the equatorial plane,
$$v_r=-v \sin \theta, ~~v_\theta=-v \cos \theta, \eqno(3.6)$$
where $v=\sqrt{(v_r^2+v_\theta^2)}$. The Mach number of the 
flow, ${\cal M}=\sqrt{(v_r^2+v_\theta^2)}/c_s$ is assumed 
to be constant and is a free parameter to be specified for a solution.
Thus, for a given polytropic constant, $K$, 
solutions of the flow can be solved from a single equation (3.1). Here,
we have assumed that ${\cal B}=0$ and $\ell=\ell_0=const$.  The vertical 
height of the flow is $H$. In addition, we have used $\gamma=5/3$.

\vskip 0.5in
\n{\bf 3.3 Torus type conditions}
\vskip 0.3in

In the case of a static torus, $v_r=0$ and $v_\theta=0$. For a given specific
angular momentum distribution, $\ell=const$, equation (3.1) can be solved 
analytically for the density distribution for a fixed ${\cal B}$ which is
a constant here. Specification of the solution is 
then determined by the polytropic constant $K$.  A general property of such 
an initial condition is that solutions cannot exist inside the funnel wall 
which, under the Newtonian potential, is determined by the condition
%%%%%%%%%%%%%%%%%%%%%%%%%%%%%%%%%%%%%%%%%%%%%%%%%%%%%%%%%%%%%%%%%%%%%%%%%%%%
$$ {\ell^2 \over 2 (r \sin \theta)^2} -{GM \over r}-{\cal B}= 0.
\eqno(3.7) $$
%%%%%%%%%%%%%%%%%%%%%%%%%%%%%%%%%%%%%%%%%%%%%%%%%%%%%%%%%%%%%%%%%%%%%%%%%%%%
In our case, the location of the funnel wall on the equatorial plane
is simply $4R_G$ for ${\cal B}=0$. It should be noted that, in the Newtonian
approximation, the funnel wall is always closed at the inner edge of the wall
independent of ${\cal B}$ and $\ell \ne 0$. 
Thus, accretion is forbidden for a static torus.

\vskip 0.5in
\n{\bf 4 Numerical Results}
\vskip 0.3in

The parameters corresponding to the numerical models are summarized in Table 1.
The initial flow type, the grid size, the outer radius, the Mach number at the 
outer edge ${\cal M}_{out}$, the two constants of $\gamma$ and ${\cal B}$,
and the vertical thickness of the initial incoming flow are listed. 
In all these models, the inner radius is fixed at $3R_G$ and the polytropic
constant is $K=4.5 \times 10^{22}$.
Also note that the Mach number is defined excluding the rotational velocity.
The Mach number is a calculated quantity in all models except in the 
Parallel-Type models in which it is a free parameter.

The numerical code was tested in two cases in which analytical solutions 
exist relevant to the proposed study. For the case of spherical Bondi 
accretion (Model~1), the exact solution is used as the initial condition.
The flow remained spherical during the entire calculation for each of the 
models tested (with varying ${\cal B}$ and $K$). A small perturbation arises
at the inner edge due to the imposed boundary condition; however, the 
perturbation propagated outwards and smoothed out very quickly (in a timescale 
much shorter than the total evolution considered). The final state is steady,
and the relative errors of the physical parameters are always less than a few 
percent.

In the second test, the evolution of a static torus (Model~2) was examined.
The analytical solution discussed above is used as the initial condition.
During the evolution, motions developed near the very surface of the torus, 
however, the motion in the interior of the torus is very small with a Mach 
number (excluding the rotational velocity) near or less than $10^{-4}$.  In 
this case, virtually no accretion occurred.

In the following, three distinct cases characterized by the initial conditions 
of the flow and the outer boundary conditions imposed are investigated.  
Specifically, they are described as follows. 

\n Case~1: The incoming flow is supersonic and is spherically distributed 
with the specific angular momentum of gas varying with the polar angle as 
$\ell(r,\theta)=\ell_0 (1-\cos \theta)$.

\n Case~2: The incoming flow is supersonic and is spherically distributed, 
but the specific angular momentum of the gas is a constant.

\n Case~3: The incoming flow is supersonic and is parallel to the equatorial 
plane. In this case the thickness of the flow is restricted and the specific 
angular momentum of gas is a constant.

Model~3 for Case~1 (see Table 1) is similar to the study first performed by 
Clarke et al. (1985).  Qualitatively, their results that accretion occurs 
primarily in the polar direction and that a shock wave forms in the inner 
region near the inner edge which then propagates outwards is confirmed. In the 
subsonic regions (behind the shock) meridional internal 
motions develop and the density distribution achieves a nearly steady pattern.  
The resolution of the present calculation is higher than in Clarke et al. 
(1985) and, hence, the circulatory motion is more apparent.  A typical flow 
pattern is illustrated in Figure 1a on a small scale ($r \lta 30 R_g$) and 
in Figure 1b on a larger scale ($r \lta 150 R_G$).  
Here, the flow vectors correspond to a quantity related to the local mass flow 
rate, $r^2 \rho \vec {v}$ (excluding the rotational velocity) and the contours 
correspond to the mass density.  At a dimensionless time of 4830 it can be 
seen in Fig. 1a that the flow is primarily radial and directed toward the 
origin for $\theta < 50$ and that several circulatory cells are present near 
the mid-plane of the flow.  From Fig. 1b, it is evident that the shock has 
propagated outward to $r \sim 105$ along the mid-plane and to $r \sim 90$ in 
the polar direction. This difference directly reflects the presence of the
centrifugal forces acting only in the direction perpendicular to the polar 
axis.  In addition, it can also be seen that there is a small 
region in the mid-plane ($50 < r < 100$) in which matter is directed away from 
the plane.   To follow the 
propagation of the shock wave, the Mach number ($v_r/c_s$) distribution (at 
the mid-plane) is plotted with respect to the radius at four different time
intervals (see Figure~2).  The shock front is well resolved and propagates 
at an average dimensionless speed of 0.15, 0.075, 0.05, and 0.025 at the four 
intervals indicated.   It is evident that the shock decelerates as a function
of time.  We note that although the initial conditions adopted for the flow 
are not identical to that in Clarke et al (1985),
the results reveal that the evolutionary pattern is very nearly the same. 

The primary result that an oblique travelling shock wave forms resulting in 
the development of vorticity behind the shock does not change even when the 
angular 
momentum distribution is constant with respect to the polar angle (Case~2). An 
example of the flow morphology is shown for Model~4 (see Fig.~3).  The flow 
in the mid-plane is similar to Model 3, but the flow away from the mid-plane 
is significantly modified.  In particular, there are two main differences in 
the flow morphology resulting from the larger amount of angular momentum in 
the flow. Namely, there is no accretion at all due to the greater contribution 
of the centrifugal force.  That is, accretion in the polar direction is absent 
inside the funnel wall.  In this case, a wind is found to form near the funnel 
wall ($r \cos \theta \sim 60$). However, due to the pressure of the incoming 
flow, the wind is relatively weak in this case. 
In fact, some of the material turns back to form a vortex-like flow
which moves outward as a function of time.
The development of a strong 
wind is favored, on the other hand, if the incoming flow 
is artificially truncated above a given vertical height.  This tendency is 
illustrated in Model~5, which is characterized by the same parameters as 
Model~4 except that the flow is truncated above $H \sim r_{out}/8$.  
It can be seen that the wind develops closer to the mid-plane at
$r \cos \theta \sim 10$ and extends to larger distances from the 
inner edge (see Fig.~4). 

The radial speed of the shock differs for each case reflecting the effect of 
the pressure of the incoming flow above the mid-plane of the disk.  
For example, 
the shock wave moves slowest in Model~5 where there is no incoming flow from 
above $H \approx 19$.  The angular momentum distribution in the incident flow 
can also affect the speed of shock propagation since the shock is found to 
travel faster in Model 4 than in Model 3.  Here, the greater centrifugal 
support
associated with the greater angular momentum in the flow acts in the same 
way to assist the enhancement of the shock propagation speed. 

The above simulations show that no standing shock waves are formed in the 
inner region of the disk.  Even in the case of parallel incident flow 
characterized by a vertical thickness of $H$, standing shock waves may not
form. In the following, we 
examine cases when a standing shock can be formed.  
Our numerical results show that the transition to this type of solution 
occurs at $H \approx 10$, or $H/r_{out} \approx 1/3$ (Model~6). 
The flow pattern for Model 6 is illustrated
in Figure 5 (at time 512) 
where it can be seen that the matter immediately behind the shock 
front is driven out as a wind away from the mid-plane.  No standing shock
is formed inside $r=31$, although it is possible that a standing
shock exists exterior to this radius.
Model~7 is corresponds to 
a boundary condition given by a parallel incident flow characterized by a 
vertical thickness of $H=8R_G$.  The flow pattern at time 4443 
is shown in Figure 6.  The sonic surface at times 4443, 2759, and 1615 is 
represented by the heavy solid, dashed, and dotted lines respectively. It can
be seen that the shock front is approximately coincident with the sonic 
surface near $r=21$ and that it is fairly steady over a long time interval. 
The shock is approximately perpendicular to the equatorial plane and extends 
in the vertical direction to about $\sim 9 R_G$.  In comparison with 
Model~6 matter in 
a larger region behind the shock is involved in the wind outflow. 

As the vertical extent of the incident flow is decreased further the position
of the "standing" shock decreases.  This is seen from Table 1 that the shock 
location is a sensitive function of $H$.  In particular, the 
radial shock position decreases from $21 R_G$ to $\sim 4 R_G$ as $H$ is 
decreased
by a factor of 2.  The flow for a very thin incident flow characterized by 
$H =4 R_G$ (Model 10) is shown in Fig. 7.  The shock is located in the very 
inner region of the flow at $r_s \sim 5$ and $r \cos \theta \sim 1 - 2$.  It 
can also be seen that the direction of the bulk flow in the wind region has 
also been affected by the height of the incident flow.  Specifically, 
the matter is ejected closer to the "disk" surface at $\theta > 60$ in 
comparison to Model~7 where the wind was not as restricted to lie as close 
to the "disk" surface (at an angle defined by $\theta > 40$). 

We point out that the shock does not reach a steady state shape and position 
as the height of the incident flow is decreased.  The shock wave in Model 10 
is found to vary in the interval $ 5R_G < r < 8R_G$ on a timescale of 2000,
which for a compact object of $10\msun$ is about 0.2 seconds.  This is 
similar to the behavior described by Ryu et al (1995) for simulations 
characterized by incident flows of small vertical extent.

The effect of numerical resolution is studied in Model 11 where
the number of grid zones are doubled in each direction. The flow
morphology is illustrated in Figure 8 where the structure in the inner region
exhibits a more well defined vortex flow. However, it can be seen from
Table 1 that the varibility of the shock position is similar to that
found in Model 10, and hence the low resolution of Model 10 provides a 
reliable description of the shock variation.

\vskip 0.5in
\n{\bf 5 Discussion and Conclusion}
\vskip 0.3in

We have systematically investigated the structure and evolution of accretion 
flows characterized by low specific angular momentum in the adiabatic
and inviscid approximation. The two dimensional hydrodynamic simulations
indicate that the accretion behavior is rich and is primarily dependent on the 
distribution of angular momentum and the vertical extent of the incoming flow.
For a specific angular momentum distribution which significantly varies 
with polar angle, accretion occurs in the polar direction only.  Accretion 
in the equatorial plane is forbidden by the centrifugal barrier
in the Newtonian approximation when the central object is located
inside the funnel wall (Ryu et al. 1995).

Shock waves are a common feature in these flows.  For an incident flow 
which is approximately spherically distributed, the shock propagates outwards
leaving behind a hot torus-like configuration.  On the other hand, for 
incident flows characterized by plane parallel geometry ($H < 10 R_G$ or 
$H/r_{out} < 1/3$) a 
steady standing shock forms.  For flows highly concentrated toward the 
equatorial plane ($H < 5 R_G$) the shock exhibits non steady behavior. 

In contrast to the incident flows described by a nearly spherical distribution,
the incident flows characterized by plane parallel conditions result in the 
ejection of matter as a weak wind emanating from the inner disk
region.  As the flow becomes more highly concentrated toward the mid-plane 
the wind increases in strength and is directed away from the disk at larger 
polar angles. 

From this study one can conclude that the travelling shocks are transitory 
features which eventually terminate at large radii.  The flow readjusts in 
this manner to the imposed initial and outer boundary 
conditions as it evolves to a quasi steady hot torus configuration.  Standing
shock fronts, on the other hand, are not transitory features in the flow. 
However, the significance of axisymmetric standing shocks in the accretion 
process may be questioned since their applicability rests on the requirement 
that the incident flow is characterized by low specific angular momentum and 
a restricted spatial vertical extent. These two restrictions are usually not 
satisfied in most astrophysical applications since geometrically thin 
accretion usually corresponds to a more nearly Keplerian flow with a high 
specific angular momentum.  Similarly, low specific angular momentum flows 
are primarily associated with geometrically thick flows which do not lead to 
standing shock structures.  Only for geometrically thin supersonic flows 
where there is insufficient time to maintain a hydrostatic balance 
in the vertical direction could such structures play a role in the accretion 
process, e.g., in the case of accretion resulting from tidal disruption
of a star close to a massive black hole.

\bigskip

\n{Acknowledgements}
\medskip

\n{X. C. thanks Phil Charles for hospitality during his visit at
Oxford University in December 1995.
This research has been supported in part by NASA under grant NAGW-2526}.
\bigskip

\newpage
\cl{\bf References}

\medskip
\refs Bisnovatyi-Kogan G. S., Kazhadan, Ya. M., Klypin, A. A., 
Lustkii, A. E., \&  Shakura, N. I. 1979, Soviet Astronomy, 23, 201
\smallskip

\refs Bondi, H. 1952, \mnras{112}, 195
\smallskip

\refs Bondi, H. \& Hoyle, F. 1944, \mnras{104}, 273
\smallskip

\refs Chakrabati, S. K. 1990, \mnras{243}, 610
\smallskip

\refs Chakrabati, S. K. \& Molteni, D. 1993, \apj{417}, 671
\smallskip

\refs Clarke, D., Karpik, S. \& Henriksen, R. N. 1985, \apjs{58}, 81
\smallskip

\refs Colella, P. \& Woodward, P. R. 1984, J. Comput. Phys. 54, 174
\smallskip

\refs Eggum, G. E., Coroniti, F. V., \& Katz, J. I. 1987, \apj{323}, 634
\smallskip

\refs Eggum, G. E., Coroniti, F. V., \& Katz, J. I. 1988, \apj{330}, 142
\smallskip

\refs Hawley, J. F., Smarr, L. L., \& Wilson, J. R. 1984a, \apj{277}, 296
\smallskip

\refs Hawley, J. F., Smarr, L. L., \&  Wilson, J. R. 1984b, \apjs{55}, 211
\smallskip

\refs Hoyle \& Lyttleton, R. A. 1939, Proc. Camb. Phil. Soc., 35, 405
\smallskip

\refs Igumenshchev, I. V., Chen, X., \& Abramowicz, M. A. 1996, 
\mnras{278}, 236
\smallskip

\refs Kley, W., \& Lin, D. N. C. 1992, \apj{397}, 600
\smallskip

\refs Molteni, D., Gerardi, G., \& Chakrabati, S. K. 1994, \apj{425}, 161
\smallskip

%\refs Papaloizou, J., \& Szuszkiewicz, E. 1994, \mnras{268}, 29
%\smallskip

%\refs Pringle, J. E. 1981, Ann. Rev. Astr. Ap., 19, 137
%\smallskip

%\refs Rees, M. J., Begelman, M. C., Blandford, R. D., \& Phinney, E. S. 1982,
%Nature, 295, 17
%\smallskip

\refs R\'o\.zyczka, M., Bodenheimer, P., \& Bell, K. R. 1994, \apj{423}, 736
\smallskip

%\refs R\'o\.zyczka, M., \& Muchotrzeb B. 1982, Acta Astron., 32, 285
%\smallskip

\refs Ruderman, M. A. \& Spiegel E. A. 1971, \apj{165}, 1
\smallskip

\refs Ruffert, M. \& Arnett, D. 1994, \apj{427}, 351
\smallskip

\refs Ryu, D., Brown, G. L., Ostriker, J. P., \& Loeb, A. 1995, \apj{452}, 364
\smallskip

\refs Shakura, N. I., \& Sunyaev, R. A. 1973, \aaa{24}, 337
\smallskip

%\refs Shapiro, S. L., Lightman, A. P., \& Eardley, D. M. 1976, \apj{204}, 178
%\smallskip

\refs Shima, E., Matsuda, T., Takeda, H., \& Sawada, K. 1985, \mnras{217}, 367
\smallskip

\refs Taam, R. E. \& Fryxell, B. A. 1988, \apj{327}, L73
\smallskip

\refs Wolfson, R. 1977, \apj{213}, 200
\smallskip

\refs Woodward, P. R. \& Colella, P. 1984, J. Comp. Phys., 54, 115
\smallskip

\clearpage
\begin{center}
\begin{tabular}{ccccccccc}
\multicolumn{9}{c}{{\bf Table 1}}\\[1.5ex]
\multicolumn{9}{c}{Steady State Models}\\[1.5ex] \hline\hline \\[-1ex]
Model & Initial Flow & $N_\theta \times N_r$ & $r_{out}$ & ${\cal M}_{out}$& 
$\gamma$ & ${\cal B}/c^2$ & $H$ &$r_{shock}$\\ [1.0ex]\hline \\[-1.5ex]
1 & Bondi $\ell=0$ & $40 \times 100$ & 149 & ... & 1.5 & ... &  & ... \\
2 & Torus $\ell=\ell_0$ & $40 \times 100$ & 149 & 0 & 1.5 & $10^{-5}$ & &...\\  
3 & case 1 & $40 \times 100$ & 149 & 3.46 & 1.5 & $10^{-5}$ &  &... \\  
4 & case 2 & $40 \times 100$ & 149 & 3.46 & 1.5 & $10^{-5}$ &  &... \\
5 & case 2 & $40 \times 100$ & 149 & 3.46 & 1.5 & $10^{-5}$ &  19 &...\\
6 & case 3 & $40 \times 60$ & 31 & 10 & 5/3 & 0 & 10 & ... \\
7 & case 3 & $40 \times 60$ & 31 & 10 & 5/3 & 0 & 8 & 21 \\
8 & case 3 & $40 \times 60$ & 31 & 10 & 5/3 & 0 & 6 & 19 \\
9 & case 3 & $40 \times 60$ & 31 & 10 & 5/3 & 0 & 5 & 7-8 \\
10 & case 3 & $40 \times 60$ & 31 & 10 & 5/3 & 0 & 4 & 5-8 \\
11 & case 3 & $80 \times 120$ & 31 & 10 & 5/3 & 0 & 4 & 5-9 

\end{tabular}
\end{center}

\clearpage
\cl{\bf FIGURE CAPTIONS}
\medskip

\n{\bf Figure 1.}
The flow pattern of model sequence 3 (see Table 1) at moment $t=4830$.
Panel (a) corresponds
to the inner region of the flow and panel (b) to the entire region of the flow.
In (a) the density ranges from $\log \rho = -9.12$ to -7.36 
($\Delta \log \rho = 0.1$)
whereas in (b) the density ranges from $\log \rho = -10.52$ to -7.56 ($\Delta 
\log \rho = 0.1$).  The vector arrow is defined as $r^2 \rho \vec {v}$, 
where $\vec {v}$ is the velocity excluding $v_\varphi$. The maximum amplitude 
of the vector 
corresponds to $4 \times 10^{13}$ g s$^{-1}$.
Note the development of vorticies in the subsonic region behind the shock 
and the accretion of matter near the polar direction.
\smallskip

\n{\bf Figure 2.}
The spatial variation of the Mach number of the flow in the equatorial plane
for model 3 for 4 different epochs.  The solid, dotted, short dashed, and 
long dashed lines correspond to 472, 1412, 2978, and 4830 respectively.  Note 
that the shock decelerates as it propagates outward.
\smallskip

\n{\bf Figure 3.}
The flow pattern of model sequence 4 (see Table 1) at moment $t=3104$.
The density ranges from 
$\log \rho = -12$ to -8.28 ($\Delta \log \rho =0.1$). 
The maximum vector corresponds
to $4 \times 10^{13}$ g s$^{-1}$. The development of 
vorticity in the inner regions is similar to that illustrated in Figure 1b 
(of model 3). Here no accretion occurs as a result of the centrifugal 
barrier.  An outflowing wind from the inner region is present.
\smallskip

\n{\bf Figure 4.}
The flow pattern of model sequence 5 (see Table 1) at moment $t=2766$. 
The density ranges from 
$\log \rho = -12$ to -8.77 ($\Delta \log \rho =0.1$). 
The maximum vector corresponds
to $4 \times 10^{13}$ g s$^{-1}$.  In this case, the 
wind is stronger than that in Figure 3.  In addition, the shock wave
moves slower.
\smallskip

\n{\bf Figure 5.}
The flow pattern of model sequence 6 (see Table 1) at moment $t=512$.
The density ranges from 
$\log \rho = -12$ to -7.22 ($\Delta \log \rho =0.1$). 
The vector arrow corresponds
to the momentum density in the flow with the largest vector equal to 100 g 
cm$^{-2}$ s$^{-1}$.  The heavy solid lines denotes a Mach number of unity.
\smallskip

\n{\bf Figure 6.}
The flow pattern of model sequence 7 (see Table 1) at moment $t=4443$.
The heavy solid, dashed, and dotted lines are the sonic surface
at moments $t=4443$, 2759, and 1615 respectively.  The shock front which is
approximately perpendicular to the equatorial plane at $r \approx 21$
is quasi-steady. Note that the vector arrow is defined as $\rho \vec {v}$ with 
a maximum of 70 g cm$^{-2}$ s$^{-1}$.  The density ranges from 
$\log \rho = -12$ to -7.03 ($\Delta \log \rho =0.1$).
\smallskip

\n{\bf Figure 7.}
The flow pattern of model sequence 10 (see Table 1) at moment $t=4161$.
The heavy solid, dashed, and dotted lines are the sonic surface
at moments $t=4161$, 3604, and 2041 respectively.  The shock front which is
approximately perpendicular to the equatorial plane at $r \approx 5-8$
is non-steady. Note that the vector arrow is defined as the momentum density
with a maximum value of 500 g cm$^{-2}$ s$^{-1}$. The density ranges from 
$\log \rho = -12.76$ to -6.62 ($\Delta \log \rho =0.1$)
\smallskip

\n{\bf Figure 8.}
The flow pattern of model sequence 11 (see Table 1) at moment $t=2288$.
The vector arrow is defined as the momentum density
with a maximum value of 700 g cm$^{-2}$ s$^{-1}$. The density ranges from 
$\log \rho = -12.0$ to -6.64 ($\Delta \log \rho =0.1$)
Note that the vortex flow in the inner region is well resolved.

\end{document}